\documentclass[5p]{elsarticle}

\usepackage{amssymb}

\journal{TIPP09 Proceedings in NIMA}

\begin{document}
\begin{frontmatter}

\title{Hadronic showers in the CALICE calorimeter prototypes}

\author[a,b]{Frank Simon\corref{cor1}}
\ead{frank.simon@universe-cluster.de}
\author{for the CALICE Collaboration}
\cortext[cor1]{Corresponding author. Tel.: +49-89-32354-535.} 
\address[a]     {Max-Planck-Institut f\"ur Physik, F\"ohringer Ring 6, 80805 M\"unchen, Germany}
\address[b]     {Excellence Cluster 'Universe', TU M\"unchen, Boltzmannstr. 2, 85748 Garching, Germany}

%
%
%
%
%
\begin{abstract}
The CALICE collaboration has constructed highly granular electromagnetic and hadronic calorimeter prototypes to evaluate technologies for the use in detector systems at the future International Linear Collider. These calorimeters have been tested extensively in particle beams at CERN and at Fermilab. We present preliminary results of an analysis of hadronic events in the combined system under test at CERN in 2006 and 2007, comprising a SiW ECAL, a scintillator tile HCAL and a scintillator strip tail catcher, the latter two with SiPM readout. The properties of hadronic showers in the HCAL, compared to simulations performed with GEANT4, are discussed. Particular emphasis is placed on the study of the linearity of the detector response and on the single particle energy resolution achievable with simple weighting algorithms based on the local energy density in the hadronic showers. 
\end{abstract}

%
%
%
%
%
%

\end{frontmatter}


%
%
%
%
%
%

\section{Introduction}

The CALICE collaboration has constructed highly granular calorimeter prototypes for experiments at the future International Linear Collider ILC. The CALICE calorimeters were tested in various different configurations in test beams at DESY, CERN and FNAL. For the results discussed here, obtained with data taken at CERN, a silicon-tungsten electromagnetic sampling calorimeter (ECAL) \cite{Anduze:2008hq}, a scintillator-steel hadron sampling calorimeter with analog readout (AHCAL) \cite{Eigen:2006eq} and a scintillator-steel tail catcher and muon tracker (TCMT) \cite{Dyshkant:2006et} were installed.

The ECAL has a total depth of 24 radiation lengths $X_0$ and consists of 30 active layers arranged in three longitudinal sections with different samplings. The first 10 layers use 1.4 mm thick tungsten absorber plates (0.4 $X_0$), followed by 10 layers of 2.8 mm thick absorbers (0.8 $X_0$) and 10 layers of 4.2 mm thickness (1.2 $X_0$). The silicon layers are segmented into individual readout pads with a size of  $1 \times 1$ cm$^2$. This results in a total of 9720 channels for the detector.

The AHCAL consists of small plastic scintillator tiles with individual readout by silicon photomultipliers (SiPMs) \cite{Bondarenko:2000in} arranged in layers between 2~cm thick stainless steel absorber plates. The size of the scintillator tiles  ranges from $3 \times 3 \ \mathrm{cm}^2$ in the center of the detector to  $12 \times 12 \ \mathrm{cm}^2$  on the outer edges of the calorimeter.  In total, the hadron calorimeter has 38 sensitive layers, amounting to a depth of 4.5 hadronic interaction lengths $\lambda_I$, with a total of 7608 scintillator cells. For the 2006 data taking period, the AHCAL was only partially instrumented with 23 active layers, leading to a coarser sampling in the back of the calorimeter.

The TCMT consists of 16 readout layers each with 20 $100 \times 5$ cm$^2$ scintillator strips read out by SiPMs between steel absorber plates, resulting in 320 readout channels. The detector is subdivided into a fine and a coarse section, where the first 8 layers have 19 mm thick absorber plates, while the absorbers for the last 8 layers are 102 mm thick.  In total, the TCMT thickness corresponds to a depth of 5.8 $\lambda_I$. 

This gives a total depth of approximately 11 $\lambda_I$ for the complete CALICE setup, and a total of 17648 readout channels. The unprecedented granularity of the CALICE calorimeters is reflected in this high number of readout channels for the test beam prototype.

In this paper we discuss preliminary results obtained with hadron beams focusing on the analog hadron calorimeter, and a study of the energy resolution of the complete setup. More results on hadronic showers in the CALICE HCAL, including studies of shower separation crucial for the performance of Particle Flow Algorithms, can be found for example in \cite{IEEE08Beni, Fabbri:2009nb}.

\section{Detector Performance and Calibration}

The analog HCAL has been operated over several month-long test beam periods. The large number of SiPMs in operation in the device is ideal for large statistics studies of the properties and performance characteristics of these devices. The pedestal distribution of all devices was monitored to detect changes of status and potential aging. For only 8 out of the almost 8000 SiPMs installed in the detector a consistent increase in the noise level with time was observed. This demonstrates that SiPMs show very stable long term behavior in the CALICE calorimeter.

The calibration of the detector was performed using a built-in LED system to calibrate the photon sensors and muon data for the cell-by-cell intercalibration, as discussed in more detail in \cite{Simon:2008qj}. Muons were also used to study the temperature behavior of the detector. Track segments identified within hadronic showers have also been used for this purpose, since they allow monitoring of the detector throughout the entire data taking periods. This technique can also be used for calibration and monitoring of the calorimeter system in a collider detector, where large number of muons are not readily available \cite{Simon:2009mw}.

Electron and positron data, taken without the electromagnetic calorimeter in front of the HCAL, were used to verify the level of detector understanding and the quality of the reproduction of detector effects in the simulation \cite{Fabbri:2009nb}. These detector simulations were performed using GEANT4 with a detailed model of the detector. The simulation results are then passed through the same reconstruction framework as the real data.

\section{Properties of Hadronic Showers in the AHCAL}

The high granularity of the AHCAL allows the study of the longitudinal and lateral profiles of hadronic showers with unprecedented three dimensional resolution. These measurements were compared with simulations using different hadronic physics lists within GEANT4 \cite{Geant4Physics}. For the results presented here, the {\verb$LHEP$} list and the GEANT4 reference list {\verb$QGSP_BERT$}, which showed best agreement with the LHC test beam data, were used, since they differ considerably in their predictions. 

\begin{figure}[hbt]
\begin{center}
\includegraphics[width=0.42\textwidth]{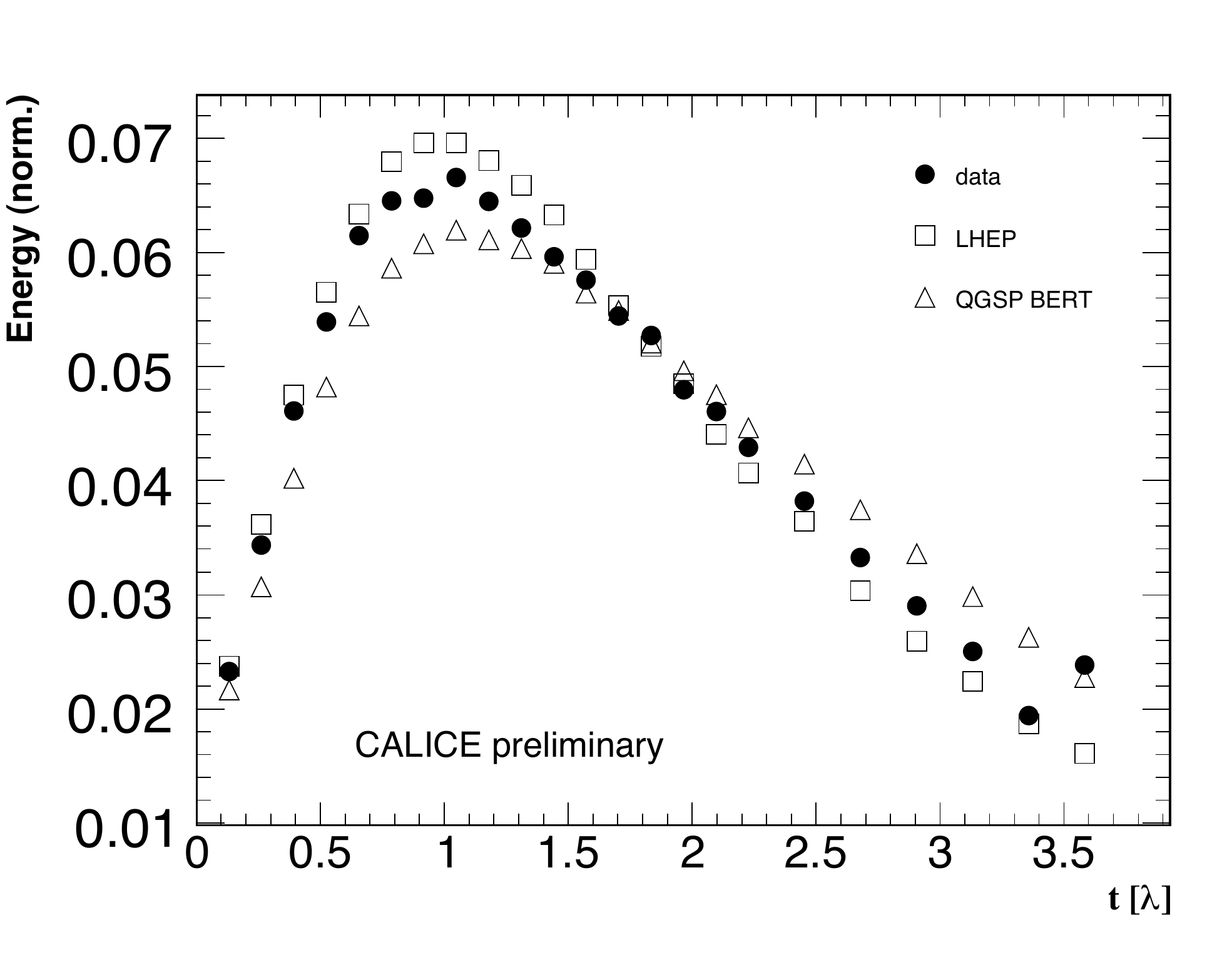}
\end{center}
\vspace{-4mm}
\caption{\label{fig:LongProfile}Longitudinal shower profile for 10 GeV negative pions, compared to simulations}
\end{figure}

Figure \ref{fig:LongProfile} shows the longitudinal shower profile for negative pions at 10 GeV. Energy deposit consistent with a minimum ionizing particle was required in the ECAL to reject events with showers starting before the HCAL. The profile was compared to simulations, which did not include a time cut to realistically mimic the integration time of the electronics. This is still under development, but was already shown to suppress the shower tails more than the core since it acts mostly on the neutron component.

The high granularity of the detector was also used to determine the starting point of each shower. For this study, an increase in hit multiplicity to  above 5 together with an increase of total signal in one layer beyond the equivalent of 8 minimum ionizing particles was taken as the start point of the hadronic cascade. To achieve increased calorimeter depth, data was taken with the hadron calorimeter rotated by 30$^\circ$, leading to a total calorimeter depth of close to 5.2 $\lambda$. In this configuration, the individual absorbers and sensitive layers are staggered so that the beam axis passes through the center of each layer.

\begin{figure}[hbt]
\begin{center}
\includegraphics[width=0.42\textwidth]{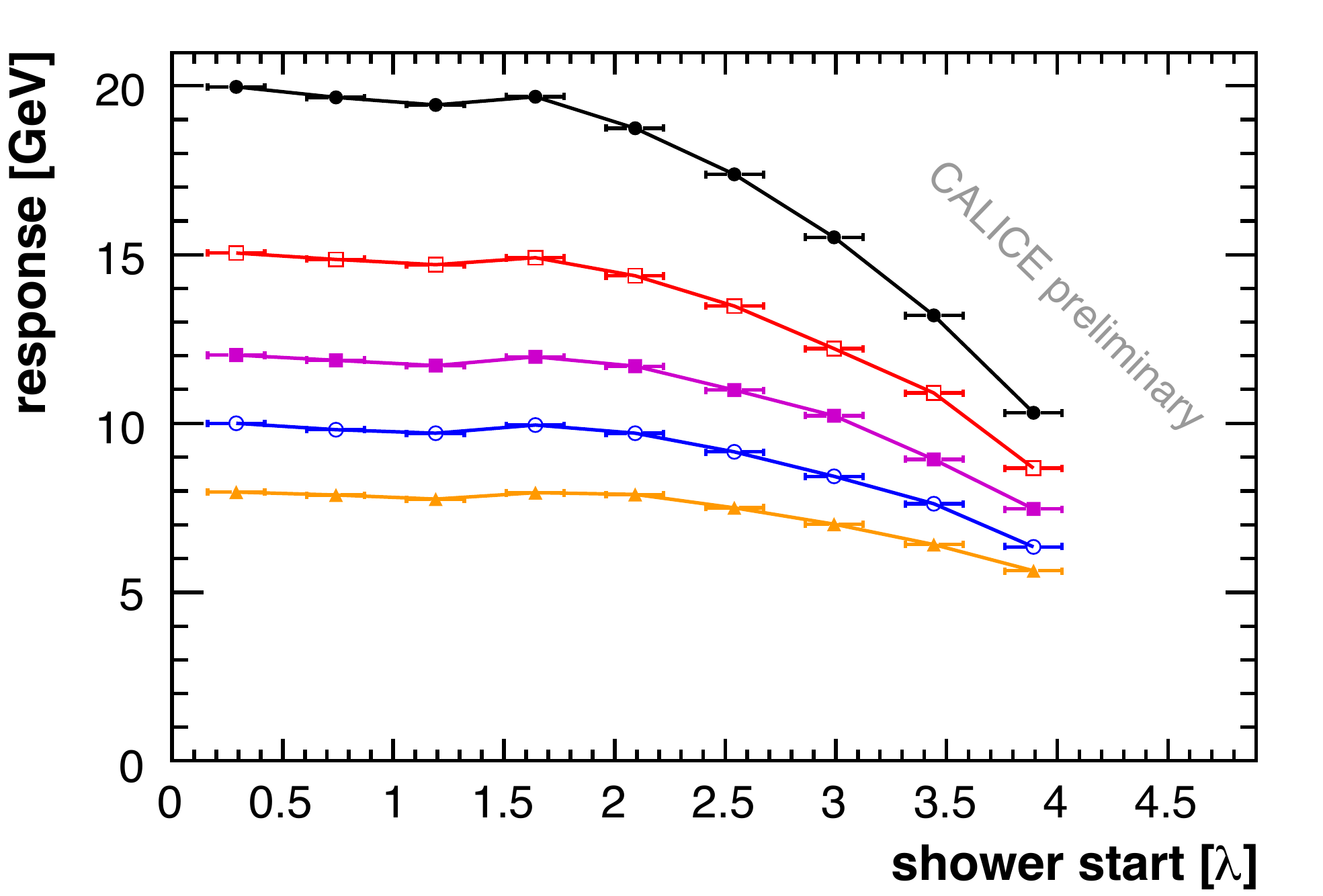}
\end{center}
\vspace{-4mm}
\caption{\label{fig:ShowerContainment}Total reconstructed energy in the HCAL as a function the shower starting point in interaction lengths.}
\end{figure}

Figure \ref{fig:ShowerContainment} shows the total reconstructed energy in the HCAL as a function of the depth of the shower start. It is clearly apparent that energy leakage is a significant effect for late starting showers, and increases with energy. The knowledge of the shower start can be used to correct the reconstructed energy. Still, the energy resolution deteriorates with increasing shower start since information is lost for events with large leakage.

\section{Energy Resolution with Shower Density Dependent Signal Weighting}

The high granularity of the CALICE detectors can also be used for software compensation procedures. For intrinsically non-compensating calorimeters like the CALICE calorimeters, the detector response is typically larger for electromagnetic than for hadronic showers. Since hadronic showers contain an electromagnetic component from the production of neutral pions in the cascade which fluctuates from event to event, this reduces the energy resolution and leads to non-linearities in the response. By identifying electromagnetic and hadronic subshowers and assigning different weights to their energy in the total reconstructed energy, resolution and linearity can be improved. Since electromagnetic showers tend to be denser than purely hadronic showers, the weights can be chosen according to the local shower density. For an initial study, each cell in a given shower was weighted with a factor chosen according to its energy content. The weights were determined from a statistically independent data set with a minimization procedure, and then parametrized to describe their energy dependence. Separate weight factors for each of the subdetectors (ECAL, HCAL, TCMT) were used. The reconstructed energy was also studied using the default case of one constant conversion factor per detector, without density dependent weighting. This reconstructed energy was also used to select the energy dependent weights, so no knowledge of the beam energy for the applications of the weighting procedure was necessary.

\begin{figure}[hbt]
\begin{center}
\includegraphics[width=0.42\textwidth]{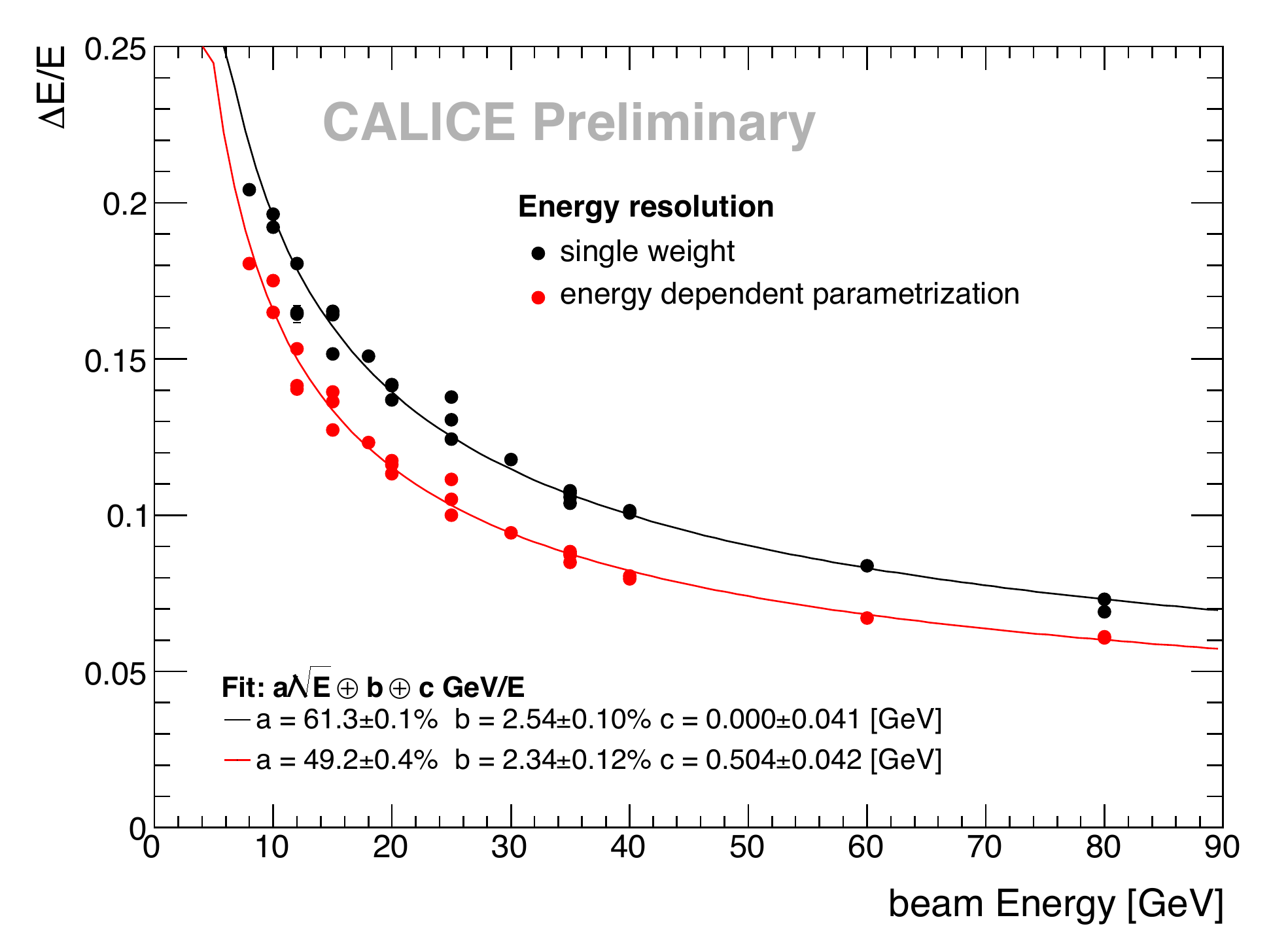}
\end{center}
\vspace{-4mm}
\caption{\label{fig:Resolution} Energy resolution of the complete CALICE setup for hadrons at various energies as a function of beam energy, using energy reconstruction with one factor converting the corrected detector signal to energy, and with local energy density dependent weighting.}
\end{figure}

Figure \ref{fig:Resolution} shows the energy resolution for the complete CALICE setup, both for the reconstruction with one conversion factor per detector and for the density dependent weights. No requirements for shower containment or shower start positions were made. It is apparent that the simple weighting method improves the energy resolution by about $20 \%$, yielding a stochastic term of $49.2\%/\sqrt{E}$.

\begin{figure}[hbt]
\begin{center}
\includegraphics[width=0.40\textwidth]{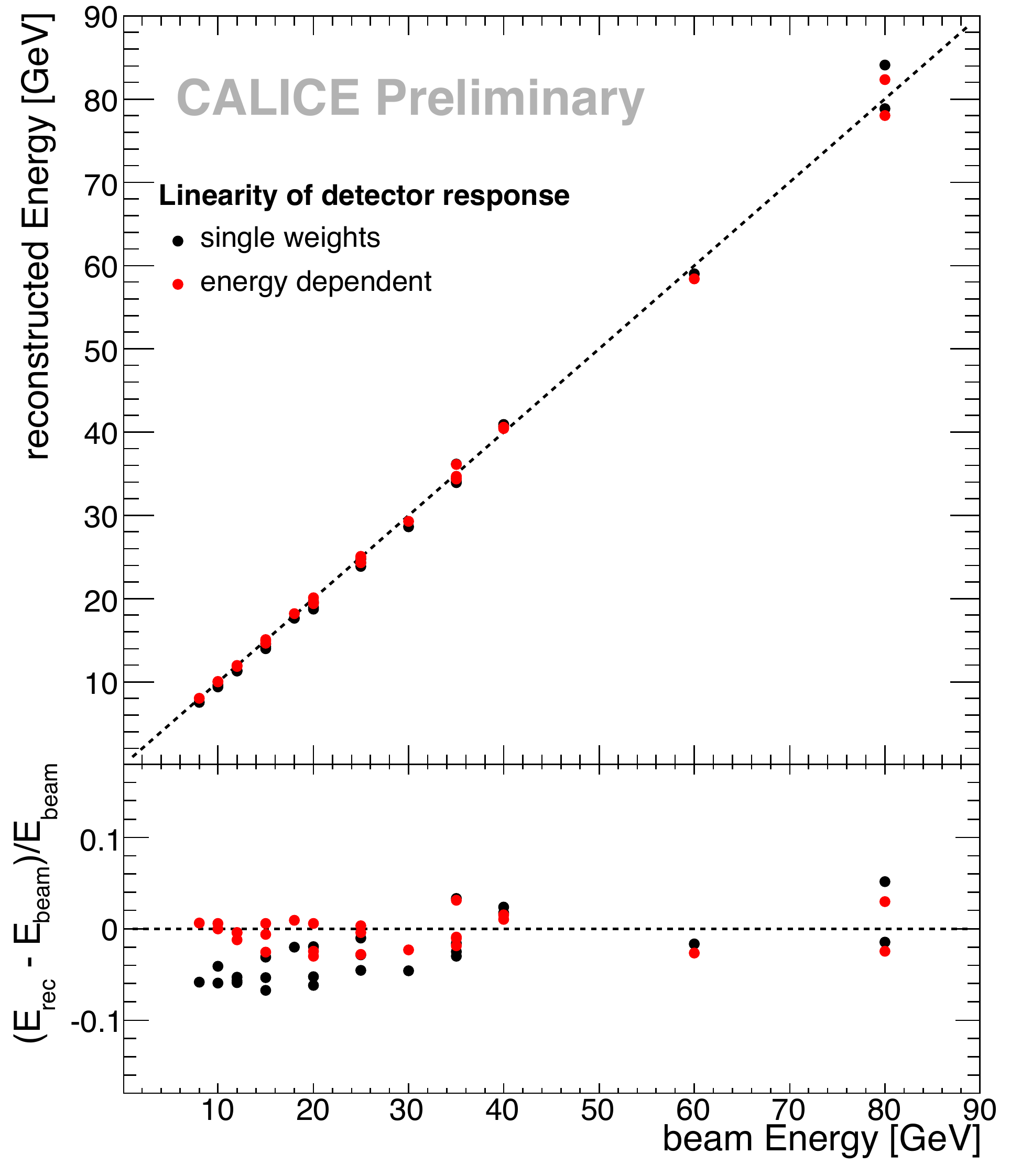}
\end{center}
\vspace{-4mm}
\caption{\label{fig:Linearity}Reconstructed energy of the complete CALICE setup for hadrons at various energies as a function of beam energy, for reconstruction with one factor and with density dependent weighting. The lower panel shows the relative deviation of the reconstructed energy from the true beam energy.}
\end{figure}

Also the linearity of the detector response was improved, as shown in Figure \ref{fig:Linearity}. The spread of reconstructed energy and resolution for multiple measurements at the same beam energy is at least partially due to missing corrections for the temperature dependence of the detector response.

\section{Summary}

The CALICE collaboration has constructed and tested calorimeters with unprecedented granularity for experiments at the future ILC. Detailed information about the properties of hadronic showers has been extracted, and was compared to GEANT4 simulations based on different physics lists. The high granularity was used to identify the start point of showers, opening up the possibility for corrections for shower leakage. Shower weighting techniques based on the local energy density were shown to improve the energy resolution significantly. Improved and expanded analyses will open up the possibility to constrain hadronic shower models in the near future with the available CALICE data set. 

%
%
%
%
%
%

\bibliographystyle{elsarticle-num.bst}

\bibliography{CALICE}

\begin{thebibliography}{1}
\expandafter\ifx\csname url\endcsname\relax
  \def\url#1{\texttt{#1}}\fi
\expandafter\ifx\csname urlprefix\endcsname\relax\def\urlprefix{URL }\fi
\expandafter\ifx\csname href\endcsname\relax
  \def\href#1#2{#2} \def\path#1{#1}\fi

\bibitem{Anduze:2008hq}
J.~Repond, et~al., {Design and Electronics Commissioning of the Physics
  Prototype of a Si-W Electromagnetic Calorimeter for the International Linear
  Collider}, JINST 3 (2008) P08001.

\bibitem{Eigen:2006eq}
G.~Eigen, {The Calice scintillator HCAL testbeam prototype}, AIP Conf. Proc.
  867 (2006) 565--573.

\bibitem{Dyshkant:2006et}
A.~Dyshkant, {Tail Catcher Muon Tracker for the CALICE test beam}, AIP Conf.
  Proc. 867 (2006) 592--599.

\bibitem{Bondarenko:2000in}
G.~Bondarenko, et~al., {Limited Geiger-mode microcell silicon photodiode: New
  results}, Nucl. Instrum. Meth. A442 (2000) 187--192.

\bibitem{IEEE08Beni}
B.~Lutz, Test beam results from the {CALICE} tile hadron calorimeter prototype
  with {SiPM} read-out, 2008 IEEE Nuclear Science Symposium Conference Record.

\bibitem{Fabbri:2009nb}
R.~Fabbri, {Tile HCAL Test Beam Analysis: Positron and Hadron Studies},
  arXiv:0902.1388 [physics.ins-det].

\bibitem{Simon:2008qj}
F.~Simon, {Calibration of a Highly Granular Hadronic Calorimeter with SiPM
  Readout}, arXiv:0811.2431 [physics.ins-det].

\bibitem{Simon:2009mw}
F.~Simon, {Track Segments in Hadronic Showers: Calibration Possibilities for a
  Highly Granular HCAL}, arXiv:0902.1879 [physics.ins-det].

\bibitem{Geant4Physics}
J.~Apostolakis, et~al., Geant4 physics lists for {HEP}, 2008 IEEE Nuclear
  Science Symposium Conference Record.

\end{thebibliography}

\end{document}